# On the Hurst Exponent, Markov Processes, and Fractional Brownian Motion

G. Millán, *Member, IEEE*

*Abstract*— There is much confusion in the literature over Hurst exponent (*H*). The purpose of this paper is to illustrate the difference between fractional Brownian motion (fBm) on the one hand and Gaussian Markov processes where *H* is different to 1/2 on the other. The difference lies in the increments, which are stationary and correlated in one case and nonstationary and uncorrelated in the other. The two- and one-point densities of fBm are constructed explicitly. The two-point density does not scale. The one-point density for a semi-infinite time interval is identical to that for a scaling Gaussian Markov process with *H* different to 1/2 over a finite time interval. We conclude that both Hurst exponents and one-point densities are inadequate for deducing the underlying dynamics from empirical data. We apply these conclusions in the end to make a focused statement about nonlinear diffusion.

## I. Introducción

La necesidad de incrementos estacionarios del movimiento browniano fraccional (fBm) ha sido enfatizada en libros [1] y artículos [2] matemáticos, pero generalmente no se menciona [3]. Los libros [4] y artículos [5] físicos tienden a ignorar por completo esta cuestión y suponer sin justificación y de manera incorrecta que $H \neq 1/2$ implica necesariamente correlaciones de largo alcance. En este artículo se enfatiza que el punto esencial en las correlaciones a largo alcance es la estacionalidad de los incrementos y no la escala. El concepto de escala hace la vida más simple pero es irrelevante para las correlaciones: existen clases de sistemas dinámicos estocásticos con correlaciones de largo alcance sin escalar, en tanto que los procesos sin memoria, procesos de Markov, pueden escalarse perfectamente con un exponente de Hurst $H \neq 1/2$ [6]. El punto es que la evidencia de escalamiento, tomada sola, no dice nada sobre la existencia de correlaciones de largo alcance. La pregunta básica que debe responderse primero tanto en análisis de datos como en teoría es: ¿son los incrementos estacionarios o no estacionarios? Para lograr la unidad y la claridad, se comienza con la definición matemática habitual de escalar un proceso estocástico y luego se demuestra cómo conduce naturalmente a la definición física.

## II. Definición del Problema

Los exponentes de Hurst son ampliamente utilizados para caracterizar procesos estocásticos, y a menudo se asocian con la existencia de autocorrelaciones que describen la memoria de largo plazo en señales. Como fue originalmente definido en [1], el exponente de Hurst (*H*) describe, entre otras cosas, el escalamiento de la varianza de un proceso estocástico, $x(t)$, a partir de la expresión

$$\sigma^2 = \int_{-\infty}^{\infty} x^2 f(x,t)dx = ct^{2H}, \qquad (1)$$

donde *c* es una constante. Aquí, el punto inicial en la serie de tiempo $x_0 = 0$ se asume como conocido en $t = 0$. Inicialmente el análisis se limita a un proceso libre de tendencia, de forma que $\langle x \rangle = 0$.

Un proceso de Markov [2], [3] es un proceso estocástico sin memoria: la densidad de probabilidad condicional para $x(t_{n+1})$ en una serie temporal $\{x(t_1), x(t_2), \ldots, x(t_n)\}$ depende solamente de $x(t_n)$ y por lo tanto es independiente de todo el historial de trayectoria anterior. La ecuación diferencial estocástica

$$dx = [D(x,t)]^{1/2} dB(t) \qquad (2)$$

genera un proceso de Markov $x(t)$, donde $B(t)$ proceso Wiener [4] con $\langle dB \rangle = 0$, $\langle dB^2 \rangle = dB^2 = dt$ ($dB/dt$ es ruido blanco).

Las soluciones de (2) son de Markov, lo cual se demuestra en [4] y [5] con detalle.

Considere a continuación lo requerido para satisfacer (1) es decir una densidad de probabilidad con un exponente de Hurst *H* tal que $0 < H < 1$, es decir

$$f(x,t) = t^{-H} F(u), \quad \text{con } u = xt^{-H}. \qquad (3)$$

La forma de escalamiento de (3) garantiza el escalamiento de la varianza en (1). Es más, de (2) también debe ser posible calcular la varianza de (1) utilizando el cálculo de Itô según la expresión

$$\sigma^2 = \int_0^t ds \int_{-\infty}^{\infty} f(x,s)D(x,s)dx, \qquad (4)$$

de forma que el coeficiente de difusión de escalamiento

$$D(x,t) = t^{2H-1}D(u), \quad u = t^{-H}x \qquad (5)$$

es también una consecuencia de (3).

Un proceso estocástico $x(t)$ escala con índice de Hurst, *H*, si satisface que

$$x(t) = t^H x(1), \qquad (6)$$

G. Millán, Facultad de Ingeniería y Tecnología, Universidad San Sebastián, Puerto Montt, Chile, ginno.millan@uss.cl.

donde la igualdad significa igualdad en las distribuciones. Lo mismo en la práctica: la distribución puntual $P(x, t)$ refleja las estadísticas recopiladas de múltiples ejecuciones diferentes de la evolución temporal de $x(t)$ a partir de una condición inicial dada $x(t_0)$, pero no describe las correlaciones o la falta de las mismas. La densidad $f_1(x, t)$ de una distribución se define por $f_1(x, t) = dP(x, t)/dx$.

Dada una variable dinámica $A(x, t)$ su promedio se calcula como sigue

$$\langle A(t) \rangle = \int_{-\infty}^{\infty} A(x,t) f_1(x,t) dx. \tag{7}$$

Se restringen los coeficientes de tendencia independientes de $x$. Sea $R(t)$ el coeficiente de tendencia de $x(t)$. Dado que la tendencia en

$$\langle x(t) \rangle = x(t_0) + \int_{t_0}^{t} R(s) ds \tag{8}$$

es dependiente de $t$, es trivial removerla del proceso estocástico eligiendo en lugar de $x(t)$ la variable martingala

$$x(t) = z(t) + \int R(s) ds \Leftrightarrow z(t) = x(t) - \int R(s) ds. \tag{9}$$

En lo que sigue se escribe $x(t)$ asumiendo que la tendencia ha sido removida, es decir $\langle x(t) \rangle = x(t_0)$, además $x(t_0) = 0$, así en términos generales $x(t)$ debe interpretarse como $x(t) - x(t_0)$ si $x(t_0) \neq 0$, es decir se usará $x(t)$ generalmente para referirse a la variable aleatoria

$$z(t) = x(t) - x(t_0) - \int \langle R dt \rangle. \tag{10}$$

De (6), los momentos de x deben obedecer

$$\langle x^n(t) \rangle = t^{nH} \langle x^n(1) \rangle = t^{nH} c_n = c_n t^{nH}. \tag{11}$$

Combinando (11) con $\langle x^n(t) \rangle = \int x^n f_1(x,t) dx$ se obtiene [6]

$$f_1(x,t) = t^{-H} F(u), \tag{12}$$

resultado compatible con (5), donde $u = x/t^H$ es la variable de escalamiento. En particular, para cumplir $\langle x(t) \rangle = x(t_0) = 0$ en un promedio sin tendencia, la varianza es de (11) simplemente

$$\sigma^2 = \langle x^2(t) \rangle = \langle x^2(1) \rangle t^{2H}. \tag{13}$$

Esto explica qué significa el exponente de escala de Hurst y determina qué significa que (6) sea en términos de igualdad de distribuciones.

En lo que sigue ecuaciones en variables aleatorias $x(t)$ como (6), por ejemplo; las soluciones de ecuaciones diferenciales de tipo estocásticas y ecuaciones incrementales como se describe en la parte III, deben entenderse como "iguales en términos de sus distribuciones". Las distribuciones de Levy no son objeto de este artículo. Para discusiones sobre el tema ver [8] y [9].

Empíricamente la mejor evidencia para el escalamiento de datos el colapso de datos dado por la forma $F(u) = t^H f_1(x, t)$ seguida por la forma, más débil, de buscar el escalamiento en un número finito de momentos $\langle x_n(t) \rangle$. Es importante entender que el escalamiento dado por el exponente de Hurst por si solo no dice nada referente a la dinámica estocástica subyacente. En particular, el escalamiento así descrito no implica ni la presencia ni la ausencia de incrementos/desplazamientos de autocorrelaciones sobre intervalos de tiempo no superpuestos, como a continuación se demuestra.

Los procesos autosimilares son fuertemente no estacionarios por (4), y los momentos no se aproximan a las constantes, sino que la densidad puntual necesariamente se extiende en ancho para siempre, reflejando así la pérdida continua de información sobre dónde se encuentra el punto $x(t)$. [6]. Pero un proceso no estacionario puede tener incrementos estacionarios o no.

III. INCREMENTOS ESTACIONARIOS VERSUS INCREMENTOS NO ESTACIONARIOS

Generalmente en la literatura se confunden los incrementos estacionarios con los procesos estacionarios (ver [6] y [7] para una discusión con todo detalle). Los incrementos estacionarios se suponen sin justificación previa en los análisis de datos por alguna razón que debe encontrarse en los autores. Se definen incrementos estacionarios y no estacionarios y demuestran sus implicancias para la cuestión de las autocorrelaciones de largo alcance o la falta completa de autocorrelaciones. Se demuestra que los incrementos estacionarios, no los incrementos de escala, son fundamentales para la existencia de correlaciones a largo plazo.

Los incrementos estacionarios $\Delta x(t, T)$ de un proceso no estacionario $x(t)$ son definidos por

$$x(t+T) - x(t) = x(T) \tag{14}$$

y por incrementos no estacionarios se requiere decir que

$$x(t+T) - x(t) \neq x(T) \tag{15}$$

Donde (14) y (15) deben entenderse en el contexto igualdad en la distribución. Las implicancias de esta distinción para el análisis de datos y para entender $H$ son centrales. Cuando (14) se cumple, debido a la densidad de posiciones $f_1(x, t)$, también se conoce la densidad $f$ de los incrementos $f(\Delta x)$,

$$f[x(t+T) - x(t)] = f_1(x, T), \tag{16}$$

lo cual facilita la construcción de lo que se quiere decir con la "igualdad en la distribución". Cuando los incrementos no son estacionarios, es imposible entonces construir una densidad de un punto $f$ de incrementos teóricamente. La razón de esto es que en el último caso la discusión de los incrementos requiere una densidad de dos puntos, $f_2(x(t), t, x(t + T), t + T)$. Por lo tanto mientras que las ecuaciones como (14) y (15) siempre se deben entender desde el punto de vista de en distribución, en el caso de (15) no existe una forma de construir una densidad de incremento independiente de $t$. Esto es particularmente cierto para los procesos de Markov donde $H \neq 1/2$.

## IV. Procesos de Markov

Se definen los procesos de Markov usando densidades y se demuestra luego que esta definición produce correlaciones nulas para incrementos sobre intervalos de tiempo no superpuestos, mostrando que los procesos de Markov no estacionarios, por lo normal generan incrementos no estacionarios.

Un proceso de Markov es un proceso estocástico sin memoria [10]-[12]: la densidad de probabilidad condicional para $x(t_n)$ en una serie temporal $\{x(t_1), x(t_2),\ldots, x(t_n)\}$ depende solo de $x(t_{n-1})$ y por lo tanto también independiente de toda trayectoria anterior $x_1, x_2, \ldots, x_{n-2}$. Para Markov la densidad de probabilidad de dos puntos $f_2(x(t_1), t_1; x(t_2), t_2)$ es suficiente. Se puede escribir

$$f_2(x(t),t;x(t+T),t+T) = g(x(t+T),t+T;x(t),t)f_1(x(t),t), \quad (17)$$

donde $f_1$ es la densidad puntual de las condiciones iniciales y $g$ es la densidad de transición o función de Green. Integrando sobre la variable anterior en $f_2$, se tiene que

$$f_1(x,t) = \int_{-\infty}^{\infty} g(x,t+T;y,t)f_1(y,t)dy. \quad (18)$$

Una condición necesaria de un proceso de Markov es

$$g(x,t;x_0,t_0) = \int_{-\infty}^{\infty} g(x,t;x',t')g(x',t';x_0,t_0)dx' \quad (19)$$

si $t_0 < t' < t$. Para el movimiento libre de tendencia suponiendo que se ha restado $\int R(t)$, se demuestra cómo la propiedad de Markov garantiza incrementos sin correlación, es decir

$$\langle (x(t_1) - x(t_1 - T_1))(x(t_1 + T_2) - x(t_2)) \rangle = 0, \quad (20)$$

sobre intervalos de tiempo no superpuestos. A priori se sabe que si no hay superposición de intervalos temporales entonces

$$[t_1 - T_1, t_1] \cap [t_2, t_2 + T_2] = \emptyset, \quad (21)$$

donde $\emptyset$ denota el conjunto vacío. Esta es una condición más débil que afirma que los incrementos son estadísticamente independientes.

En primer lugar, tenga en cuenta que si los incrementos no son estacionarios, incluso si conocemos la función de Green $g$, no conocemos la densidad correspondiente a los *incrementos*. Para probar que un proceso de Markov garantiza incrementos sin correlación, se puede formular el problema como sigue. Se puede probar la falta de autocorrelaciones $\langle (x(t_1) - x(t_1 - T)(t_2 + T) - x(t_2)) \rangle$ para intervalos de tiempo no superpuestos $t_1 < t_2$ con $T > 0$, si se demuestra que las autocorrelaciones $\langle x(t)x(t + T) \rangle$ se reducen al segundo momento en el menor tiempo. Es decir, con $T > 0$ en

$$\langle x(t)x(t+T) \rangle = \iint xyg(y,t+T;x,t)f_1(x,t)dxdy, \quad (22)$$

se debe mostrar que $\langle x(t)x(t + T) \rangle = \langle x^2(t) \rangle$ si $T > 0$.

Se verifica que lo anterior es cierto para una martingala [12], porque el promedio condicional de $x(t + T)$ comenzando desde un punto $x(t)$ es

$$\int yg(y,t+T;x,t)dy = x, \quad (23)$$

con rendimientos dados por

$$\langle x(t)x(t+T) \rangle = \iint x^2 f_1(x,t)dx = \langle x^2(t) \rangle, \text{ for } T > 0. \quad (24)$$

Si se trabaja con la martingala $z(t) = x(t) - \int R(s)ds$, la cual es variable, en lugar de hacerlo con $x(t)$, entonces se prueba la falta de autocorrelaciones incrementales.